\documentclass[aps,prc,twocolumn,showpacs,nofootinbib]{revtex4-1}
\usepackage[utf8]{inputenc}
\usepackage{graphicx}   
\usepackage{latexsym}   
\usepackage{enumerate}
\usepackage{rotating,booktabs,multirow}
\usepackage{amsmath}
\usepackage{amsfonts}
\usepackage{amssymb}
\usepackage{multirow}
\usepackage{bm}
\usepackage{xcolor}

\begin{document}
\title{Correcting the $B_A$ coalescence factor at GSI-HADES and RHIC-BES energies}
\author{Apiwit Kittiratpattana$^{1,4}$, Tom Reichert$^{1,3}$, Jan~Steinheimer$^{5}$, Christoph Herold$^4$, Ayut Limphirat$^4$, Yupeng Yan$^4$, Marcus~Bleicher$^{1,2,3}$}

\affiliation{$^1$ Institut f\"ur Theoretische Physik, Goethe Universit\"at Frankfurt, Max-von-Laue-Strasse 1, D-60438 Frankfurt am Main, Germany}
\affiliation{$^2$ GSI Helmholtzzentrum f\"ur Schwerionenforschung GmbH, Planckstr. 1, 64291 Darmstadt, Germany}
\affiliation{$^3$ Helmholtz Research Academy Hesse for FAIR (HFHF), GSI Helmholtz Center for Heavy Ion Physics, Campus Frankfurt, Max-von-Laue-Str. 12, 60438 Frankfurt, Germany}
\affiliation{$^4$ Center of Excellence in High Energy Physics \& Astrophysics, School of Physics, Suranaree University of Technology, University Avenue 111, Nakhon Ratchasima 30000, Thailand}
\affiliation{$^5$ Frankfurt Institute for Advanced Studies (FIAS), Ruth-Moufang-Str.1, D-60438 Frankfurt am Main, Germany}

\begin{abstract}
We investigate the coalescence factors $B_2$ and $B_3$ at low collision energies ($\sqrt{s_\mathrm{NN}}<6$ GeV) with special focus on the HADES and RHIC-BES experiments. It is shown that, in order to properly interpret the coalescence factors $B_A$, two important corrections are necessary: I) $B_2$ has to be calculated using the proton $\times$ neutron yields in the denominator, instead of the square of the proton yield, and II) the primordial proton (neutron) densities have to be used for the normalization and not the final state (free) protons (neutrons). Both effects lead to a drastic reduction of $B_2$ and $B_3$ at low energies.
This reduction decreases the discrepancy between the volumes extracted from HBT measurements and the volumes extracted from the coalescence factor ($V\propto 1/B_2$). While at HADES and low RHIC-BES energies these corrections are substantial, they become irrelevant above $\sqrt{s_\mathrm{NN}}>6$ GeV. The proposed correction method is model independent and is only based on the measurement of protons, clusters and charged pions. 
\end{abstract}

\maketitle

\section{Introduction}
The production of nuclear clusters has again become a field of active research in heavy ion physics over the last ten years \cite{Andronic:2010qu,Sun:2018jhg,Braun-Munzinger:2018hat}. These activities have mainly been triggered by the observation of the Lambda-hypertriton at LHC and also by the recent measurements of clusters and anti-clusters up to Helium (anti-nuclei) at RHIC and LHC \cite{STAR:2011eej,ALICE:2017jmf}. These measurements have not only been interesting for the heavy ion community but are also relevant for our understanding of cosmological models of Dark Matter \cite{Bellini:2022srx,Serksnyte:2022onw}. Various (partly controversial) models and ideas for the calculation of nuclear clusters have been put forward and discussed in the literature. These models can be separated broadly into two categories: I) thermal models, treating the clusters as entities in the ensemble \cite{Bebie:1991ij,Braun-Munzinger:1994ewq,Braun-Munzinger:1995uec,Andronic:2017pug,Xu:2018jff,Vovchenko:2019aoz,Neidig:2021bal} and II) coalescence models, creating the clusters from the primordial (frozen-out) nucleons, if the nucleons are sufficiently close in phase-space \cite{Aichelin:1991xy,Nagle:1994wj,Bleicher:1995dw,Nagle:1996vp,Puri:1996qv,Puri:1998te,Ko:2010zza,Botvina:2014lga,Zhu:2015voa,Botvina:2016wko,Sombun:2018yqh,Sun:2020uoj,Zhao:2021dka,Kireyeu:2022qmv,Kittiratpattana:2020daw}. 

\begin{figure}
    \centering
    \includegraphics[width=\columnwidth]{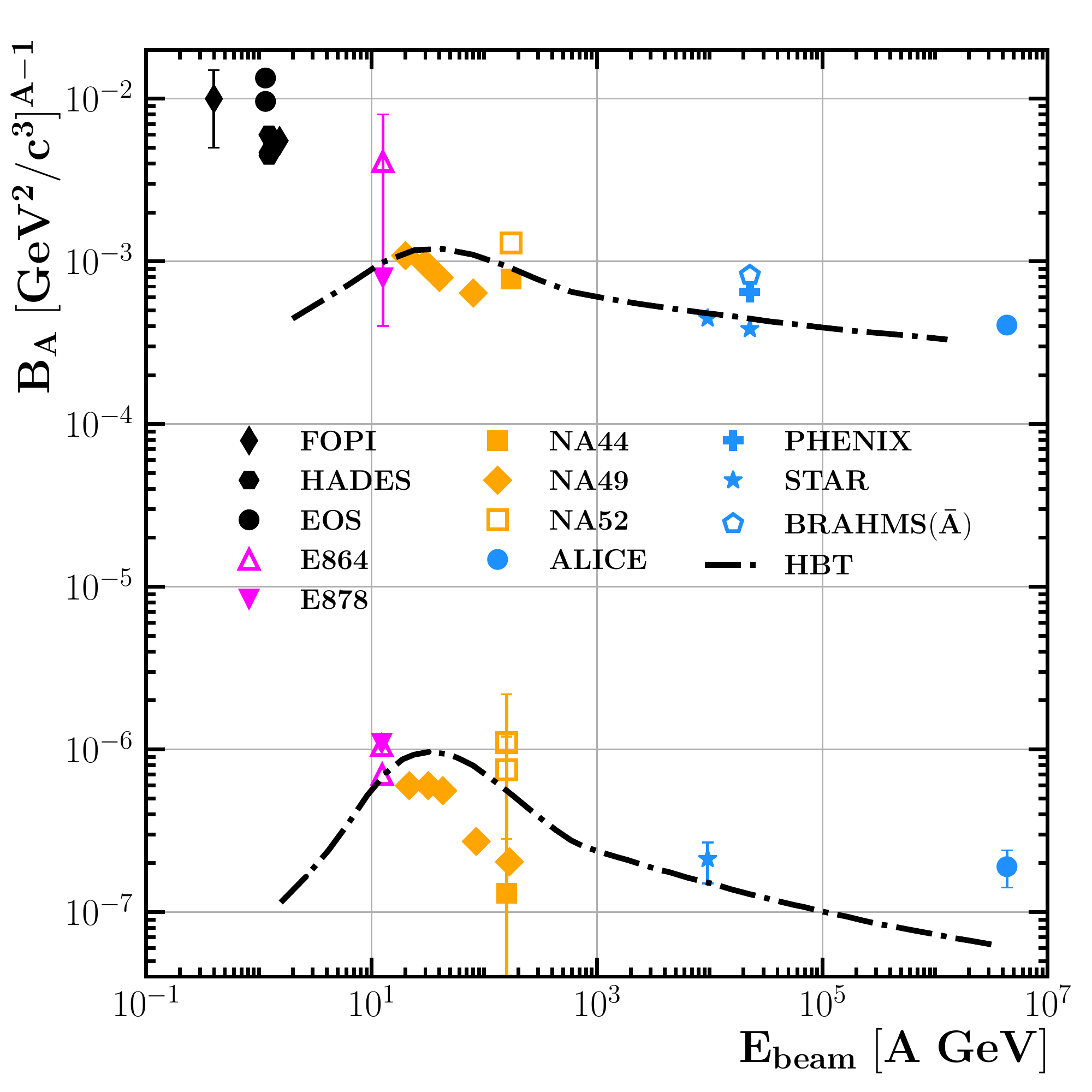}
    \caption{[Color online] Coalescence factors $B_2$ (top) and $B_3$ (bottom) for nuclear collisions as a function of beam energy. The symbols are the $B_A$ values extracted by various experiments, the lines show the scaling expected from the HBT measurements from STAR \cite{STAR:2014shf}. Data taken from \cite{Braun-Munzinger:2018hat,FOPI-B2,EOS:1994jzn,HADES-B2}.}
    \label{fig:BA_sqrts}
\end{figure}

Already early on, Ref. \cite{Butler:1963pp} suggested that the deuteron number in a given momentum interval should be proportional to the product of the proton and neutron numbers at half the deuteron momentum\footnote{Let us note that the main assumption of the model of Butler and Pearson \cite{Butler:1963pp} is that the coalescing nucleons are homogeneously distributed in the volume. Such an assumption is often not justified and has led to the usage of phase space coalescence models. Nevertheless, the suggested scaling relation between nucleons and clusters is often used to interpret experimental data, because it only relies on momentum space information.}. This equation was then extended to higher mass clusters and is often used for the interpretation of the cluster yield. Quantitatively, the relation between the density of nucleons and the final cluster yield is then:
\begin{equation}
    E_A\frac{{\rm d}^3N_A}{{\rm d} p_A^3} = B_A \left(E_p\frac{{\rm d}^3N_p}{{\rm d} p_p^3}\right)^Z \left(E_n\frac{{\rm d}^3N_n}{{\rm d} p_n^3}\right)^N, \label{eq:coal}
\end{equation}
where the evaluation is done at the same momentum per nucleon. Here, $A=N+Z$ is the mass number of the produced cluster with neutron number $N$ and proton number $Z$ and $B_A$ is the so called coalescence factor. The coalescence factor can be related to the spatial volume $V$ of the source via $B_A \propto (1/V)^{A-1}$. Such a scaling is obvious when one assumes thermal sources, but is also fulfilled in the coalescence picture, see e.g. \cite{Kapusta:1980zz,Nagle:1996vp}, however, with slightly different numerical pre-factors. The exact interpretation of the ``volume'' depends on the model. In a thermal model, it is the volume of the thermal source, in coalescence approaches it is typically associated with the volume out of which the clusters coalesce \cite{Kapusta:1980zz}, similar to the region of homogeneity in Hanbury-Brown-Twiss (HBT) analyses. It is now tempting to use the source volume (region of homogeneity) extracted via HBT correlations to check whether the $B_A$ values from cluster measurements are compatible (up to an overall scaling factor) with the HBT measurements as a function of energy. This comparison is shown in Fig. \ref{fig:BA_sqrts} (where the data is taken from \cite{Braun-Munzinger:2018hat,FOPI-B2,EOS:1994jzn,HADES-B2}). Here one observes that the $B_A$ values do indeed surprisingly well scale with the (inverse) HBT-volume down to beam energies of around $10 A$\ GeV (or center-of-mass energies around $\sqrt{s_\mathrm{NN}}=5$ GeV). At lower collision energies one observes a strong discrepancy between the $B_2$ measurements and the HBT measurements: The HBT measurements suggest a decrease of $B_2$, while the experimental data seems to suggest a strong increase of $B_2$. Unfortunately, there is no data on $B_3$ at low collision energies, however taking the ratios of the integrated yields does also imply a similar increase of $B_3$ at low energies.

In this paper, we want to address this discrepancy at low collision energies and clarify the reasons for this mismatch. We will conclude that the discrepancy can be resolved if the measured data is properly corrected.

\section{The problem with $B_A$}
Let us shortly remind the reader of the meaning of Eq. \eqref{eq:coal}. This equation describes the formation of a cluster of mass $A$ from the given primordial densities of protons and neutrons. With primordial, we mean the protons and neutrons at freeze-out before the clusters have been formed. However, there are two problems when using Eq. \eqref{eq:coal} to analyze experimental data at low energies:
\begin{itemize}
     \item Only the final state distributions of free protons (and estimated neutrons) are used to infer $B_A$, while Eq. \eqref{eq:coal} contains all protons and neutrons, including also those which will eventually end up in a cluster.  
    \item The densities of neutrons are usually not measured and the neutron distribution is estimated to be the same as the proton distribution.
\end{itemize}

At high collision energies both assumptions are well justified, because the yield of clusters is very small compared to the yields of protons and neutrons and also the yield of neutrons approaches the yield of protons due to copious pion production. However, towards lower collision energies both assumptions are not satisfied any longer. 

At low collision energies, as e.g., explored by the FOPI and HADES experiments at GSI, and at the lowest RHIC-BES energy, one can not assume that there is sufficient pion production to lead to an equipartition of isospin between neutrons and protons. In contrast, one should expect that the initial isospin asymmetry prevails to a certain extent, i.e., $N_n/N_p \approx N_{\rm Au}/Z_{\rm Au}=1.49$. This means that the proton number is not a good proxy for the neutron number. In addition, a large number of clusters, ranging from deuterons and tritons to Heliums are produced. In fact, approximately 40\% of the protons at midrapidity are captured in clusters, thus the assumption that the final state protons (neutrons) are a good proxy of the primordial protons (neutrons) is also not justified.

In the following, we will discuss how to properly treat these effects to obtain $B_A$ values that can be compared in a meaningful way to high energy data and to the HBT-volume. To this aim, we will use the UrQMD transport model to quantify the magnitude of these effects. However, the methodology is model independent and can be directly applied to experimental data.

\section{The UrQMD transport approach}
For the present study we use the Ultra-relativistic Quantum Molecular Dynamics (UrQMD) model \cite{Bass:1998ca,Bleicher:1999xi,Bleicher:2022kcu} in its most recent version (v3.5). UrQMD is a dynamical microscopic transport simulation based on the explicit propagation of hadrons in phase-space. The imaginary part of the interactions is modeled via binary elastic and inelastic collisions, leading to resonance excitations and decays or color flux-tube formation and their fragmentation. In its current version, UrQMD includes a large body of baryonic and mesonic resonances up to masses of 4~GeV. The real part of the interaction potential is implemented via different equations of state (following the usual notion of a hard and soft Skyrme-type equation of state). But alternative equations-of-state, including a chiral mean field EoS, see e.g. \cite{Kuttan:2022zno} can also be employed. For the present study, we use UrQMD in cascade mode. The model is well established and tested in the GSI/RHIC-BES energy regime. For recent studies of the bulk dynamics, we refer the reader to \cite{Reichert:2020uxs,Reichert:2021ljd}. For the details and previous results of the phase space coalescence approach used for the present study, we refer the reader to \cite{Sombun:2018yqh,Gaebel:2020wid,Hillmann:2021zgj,Kireyeu:2022qmv}. The light clusters which are included in this analysis are the deuteron ($d$), the triton ($t$) and the Helium-3 ($^3$He).

\section{Results}
We will focus the main discussion on central ($\sigma/\sigma_\mathrm{tot}=10\%$, corresponding to $b\leq 4.7$ fm in the simulations) Au+Au reaction at $E_\mathrm{beam}=1.23A$\ GeV. For this system the HADES experiment has recently measured clusters and estimated preliminary $B_2$ values, which allow to demonstrate the discussed effects. We will use this example to introduce the corrections to $B_2$ (and $B_A$ in general) and employ the same correction techniques to further energies to show until which energy the corrections are important and how they modify the $B_A$ values.  

\subsection{Primordial protons vs. final state protons}
\begin{figure}
    \centering
    \includegraphics[width=\columnwidth]{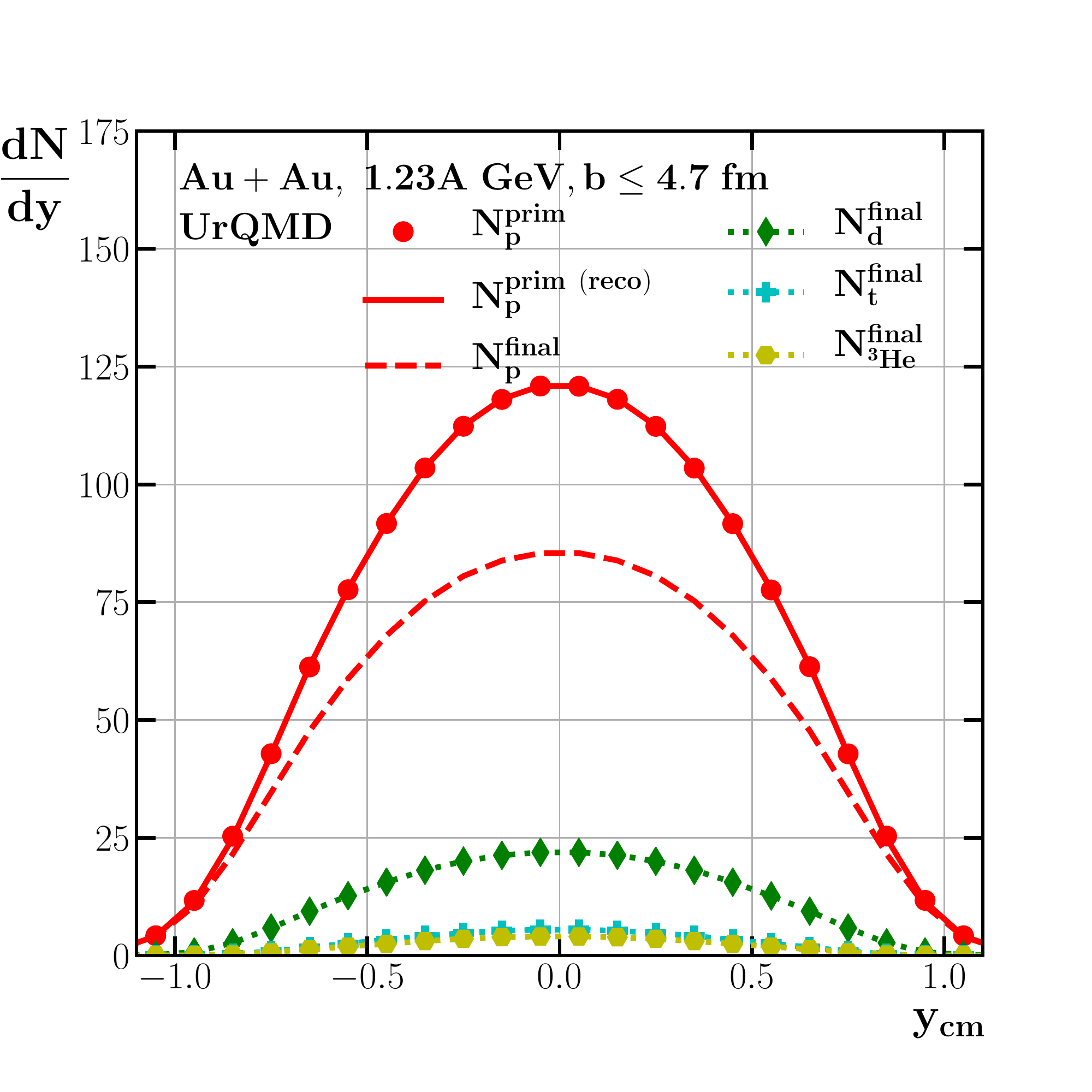}
     \caption{[Color online] Rapidity distributions of protons and light nuclei in central Au+Au reactions at ${E_\mathrm{beam}=1.23A\ \rm GeV}$. From bottom to top we show the distributions of final state ${}^3$He (yellow hexagons, dotted line), tritons (cyan pluses, dotted line), deuterons (green diamonds, dotted line) and protons (dashed red line). We compare the reconstructed primordial protons (solid red line) to the simulated primordial proton yield (full red circles).}
    \label{fig:dndy_proton_initial_vs_final}
\end{figure}

First we explore the difference between the final state (measured) proton distributions and the primordial (before clustering) proton distributions in the simulation. This requirement is especially important at very low energies discussed here in relation to the currently operating HADES experiment due to the rather large fraction of light clusters.
Fig. \ref{fig:dndy_proton_initial_vs_final} shows the rapidity distribution in central Au+Au reactions at $E_\mathrm{beam}=1.23A$\ GeV. From bottom to top we show the distributions of final state $3$-Helium (yellow hexagons, dotted line), tritons (cyan pluses, dotted line), deuterons (green diamonds, dotted line) and protons (dashed red line). We then compare the final state protons (dashed red line) with the simulated primordial proton yield (full red circles). One clearly sees that the final state proton yields are substantially lower than the primordial proton yields. Adding the protons from the clusters to the final state protons using Eq. \eqref{eq:proton_reco} allows to reconstruct the primordial proton distribution with high accuracy (the reconstructed proton distribution is shown by the solid red line to be compared to the full red squares).

For the reconstruction, we suggest to use the individual rapidity densities and the following equation:
\begin{equation}
    \frac{{\rm d}N_p^{\rm prim\,(reco)}}{{\rm d}y} = \frac{{\rm d}N_p^{\rm final}}{{\rm d}y} + \sum\limits_{\rm cluster}Z_c^p\frac{\mathrm{d}N_c^{\rm final}}{\mathrm{d}y} \label{eq:proton_reco}
\end{equation}
Here ${\rm d}N_p^{\rm prim\,(reco)}/{\rm d}y$ is the reconstructed proton distribution, while ${\rm d}N_p^{\rm final}/{\rm d}y$ is the final state rapidity distribution, ${\rm d}N_c^{\rm final}/{\rm d}y$ is the final state rapidity distribution of the cluster $c$, and $Z_c^p$ is the proton number of the respective cluster.
Thus, the primordial proton distribution can be directly reconstructed by adding the measured distributions of the clusters with their respective proton number in each rapidity bin to the final proton yields. Quantitatively, at HADES energies, only 60\% of the protons are observed as free protons in the final state, while the rest is bound in cluster states. Thus, already the use of the primordial proton densities (i.e. before clustering) instead of the final ones leads to a reduction of $B_2$. We note that clusters with larger mass numbers, especially the $^4$He, would also need to be taken into account to reliably measure $B_2$ value in the experiment. The $^4$He yield in this energy range is expected to be between 2-4, in line with FOPI measurements \cite{FOPI:2010xrt} and multi-fragmentation models \cite{Botvina:2020yfw}. 

After the reconstruction of the primordial proton distribution, we need to estimate the primordial neutron distribution from the measured spectra.

\subsection{Estimating the neutron distribution from the proton distribution}
To estimate the primordial neutron distribution we start from the initial state of the reaction. Before their collision, each isospin of the gold nuclei is distributed asymmetrically, i.e. $N_{\rm Au}/Z_{\rm Au}=1.49$. However, towards the final state of the reaction, the isospin becomes more equally distributed among the baryons due to the emission of charged pions. I.e., during the evolution (at such low energies) the emission of a $\pi^+$ goes together (after the decays of intermediate states) with the transformation of a proton into a neutron, while the emission of a $\pi^-$ is tied to a neutron being transformed into a proton. This implies that the proton and neutron numbers integrated over all participants before coalescence at a given centrality (i.e. fixed $A_{\rm part}$) can be related to the initial proton and neutron numbers via $N_p^{\rm prim\,(reco)} = \frac{A_{\rm part}}{A}Z_{\rm Au} + (N_{\pi^-} - N_{\pi^+})$ and $N_n^{\rm prim\,(reco)} = \frac{A_{\rm part}}{A}N_{\rm Au} - (N_{\pi^-} - N_{\pi^+})$. Putting both equations together and by abbreviating $\Delta\pi \equiv N_{\pi^-} - N_{\pi^+}$, this provides a good estimate for the reconstructed primordial neutron number as a function of the primordial proton number and the charged pion yield:
\begin{equation}\label{eq:n_beforec_with_Apart}
    N_n^{\rm prim\,(reco)} = N_p^{\rm prim\,(reco)} \frac{\frac{A_{\rm part}}{A}N_{\rm Au} - \Delta\pi}{\frac{A_{\rm part}}{A}Z_{\rm Au} + \Delta\pi}\quad.
\end{equation}
Here, the number of primordial neutrons is expressed in terms of the pion numbers and the primordial proton number which can be measured. However, the number of participating protons and neutrons (before isospin equilibration) cannot be measured directly and has to be inferred. It is possible to obtain a closed formula for the number of participants by making use of the relation $N_n^{\rm part} = \frac{N_{\rm Au}}{Z_{\rm Au}}N_p^{\rm part}$ and by inserting $N_p^{\rm part} = N_p^{\rm prim} - \Delta\pi$ which quantifies the amount of isospin equilibration. By requiring that $A_{\rm part} = N_p^{\rm part} + N_n^{\rm part}$, this yields in Eq. \eqref{eq:Apart}.
\begin{equation}\label{eq:Apart}
A_{\rm part} = (N_p^{\rm prim}-\Delta \pi)\times \left(\frac{N_{\rm Au}}{Z_{\rm Au}}+1\right)
\end{equation}
This result can be reinserted into Eq. \eqref{eq:n_beforec_with_Apart} in order to obtain a closed formula for the number of primordial neutrons.
\begin{widetext}
\begin{equation}\label{eq:n_beforec}
    N_n^{\rm prim\,(reco)} = N_p^{\rm prim\,(reco)} \underbrace{\frac{(N_p^{\rm prim}-\Delta \pi) \left(\frac{N_{\rm Au}}{Z_{\rm Au}}+1\right) \frac{N_{\rm Au}}{A} - \Delta\pi}{(N_p^{\rm prim}-\Delta \pi) \left(\frac{N_{\rm Au}}{Z_{\rm Au}}+1\right) \frac{Z_{\rm Au}}{A} + \Delta\pi}}_{\equiv \delta_{\rm iso}^{\rm prim}}\quad.
\end{equation}
\end{widetext}

It is obvious that this relation holds for all participants in $4 \pi$, however it is not a priori clear that this relation also holds for the differential distributions, because the width of the rapidity distributions of protons/clusters in comparison to the pions are not the same. We have therefore explored both options: I) a momentum differential isospin correction factor and II) an integrated isospin correction factor. We found that the use of the integrated correction factor provides better results for the reconstruction of the primordial neutron distributions. We abbreviate the isospin factor as $\delta_{\rm iso}^{\rm prim}$ which relates the primordial neutrons to the primordial protons\footnote{Note that this factor is energy dependent. At 1.23$A$ GeV it evaluates to the numerical value of $\delta_{\rm iso}^{\rm prim}=1.32$ which is somewhat between the initial value of 1.49 and 1 (complete isospin equilibration).}.

The reconstructed primordial neutron distribution can now be expressed via the measured rapidity distributions of protons and clusters and the integrated number of charged pions. The primordial neutron rapidity distribution is then given by:
\begin{align}\label{eq:N_n initial}
    \frac{{\rm d}N_n^{\rm prim\,(reco)}}{{\rm d}y} &= \left(\frac{{\rm d}N^{\rm final}_p}{{\rm d}y} + \sum\limits_{\rm cluster}Z_c^p\frac{\mathrm{d}N_c^{\rm final}}{\mathrm{d}y} 
    \right) \delta_{\rm iso}^{\rm prim}.
\end{align}

\begin{figure}
    \centering
    \includegraphics[width=\columnwidth]{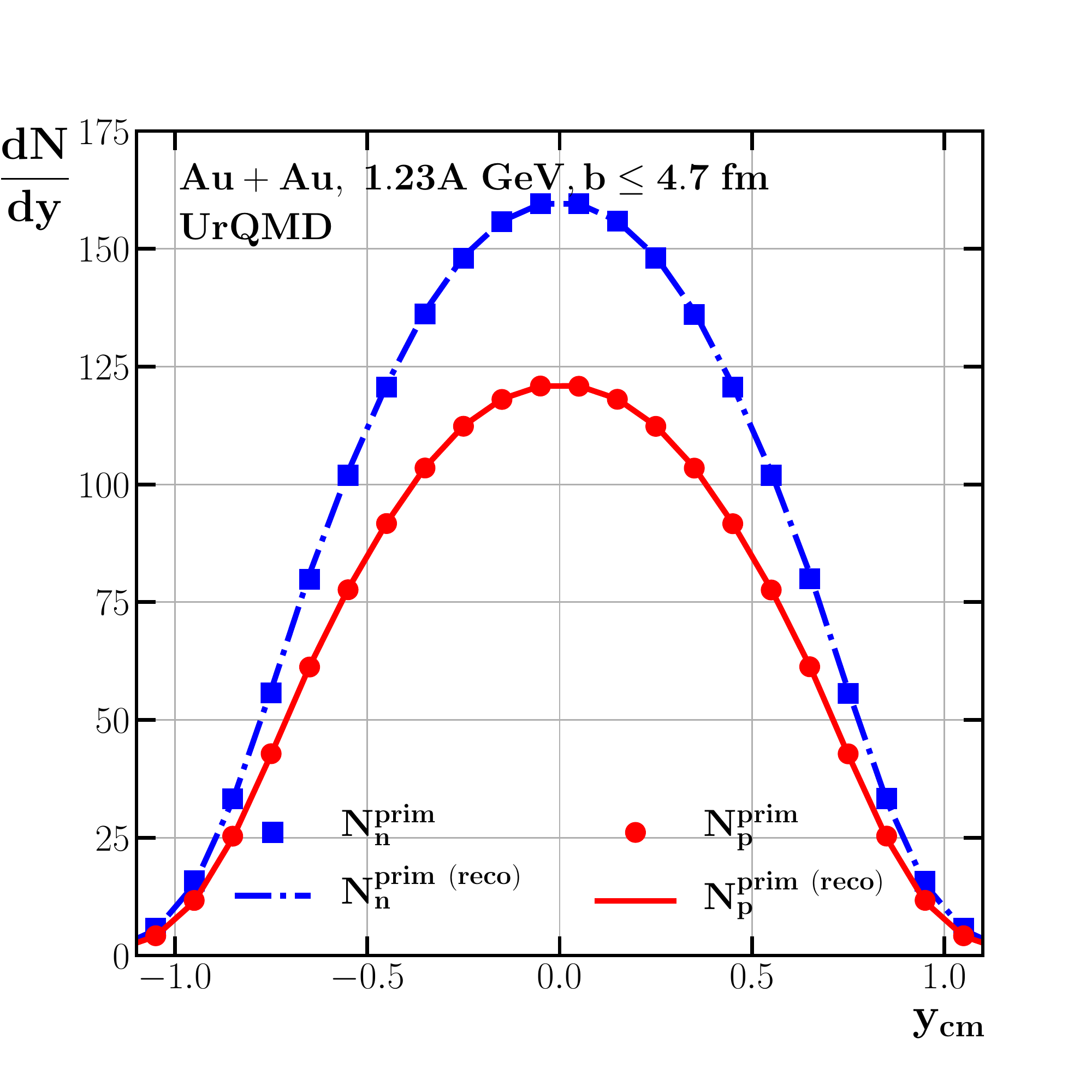}
    \caption{[Color online] Rapidity distribution of primordial protons (simulated: full red circles, reconstructed: solid red line) and our estimate for the neutron distribution based on Eq. \eqref{eq:N_n final} (dash-dotted blue line) in comparison to the neutron distribution from the simulation (full blue squares) in central Au+Au reactions at $E_\mathrm{beam}=1.23A$\ GeV.}
    \label{fig:dndy_neutrons_estimator}
\end{figure}

To quantify the quality of our reconstruction of the primordial neutron distribution, we show in Fig. \ref{fig:dndy_neutrons_estimator} the distribution of primordial protons (simulated: full red circles, reconstructed: solid red line) and our distribution of the neutron rapidity density (simulated: full blue squares, reconstructed using Eq. \eqref{eq:N_n initial}: dash-dotted blue line). Comparing the blue squares with the dash-dotted blue line, we observe a remarkably good quality of the reconstruction of the primordial neutron density, thus validating our suggestion for the reconstruction method.

\begin{figure}
    \centering
    \includegraphics[width=\columnwidth]{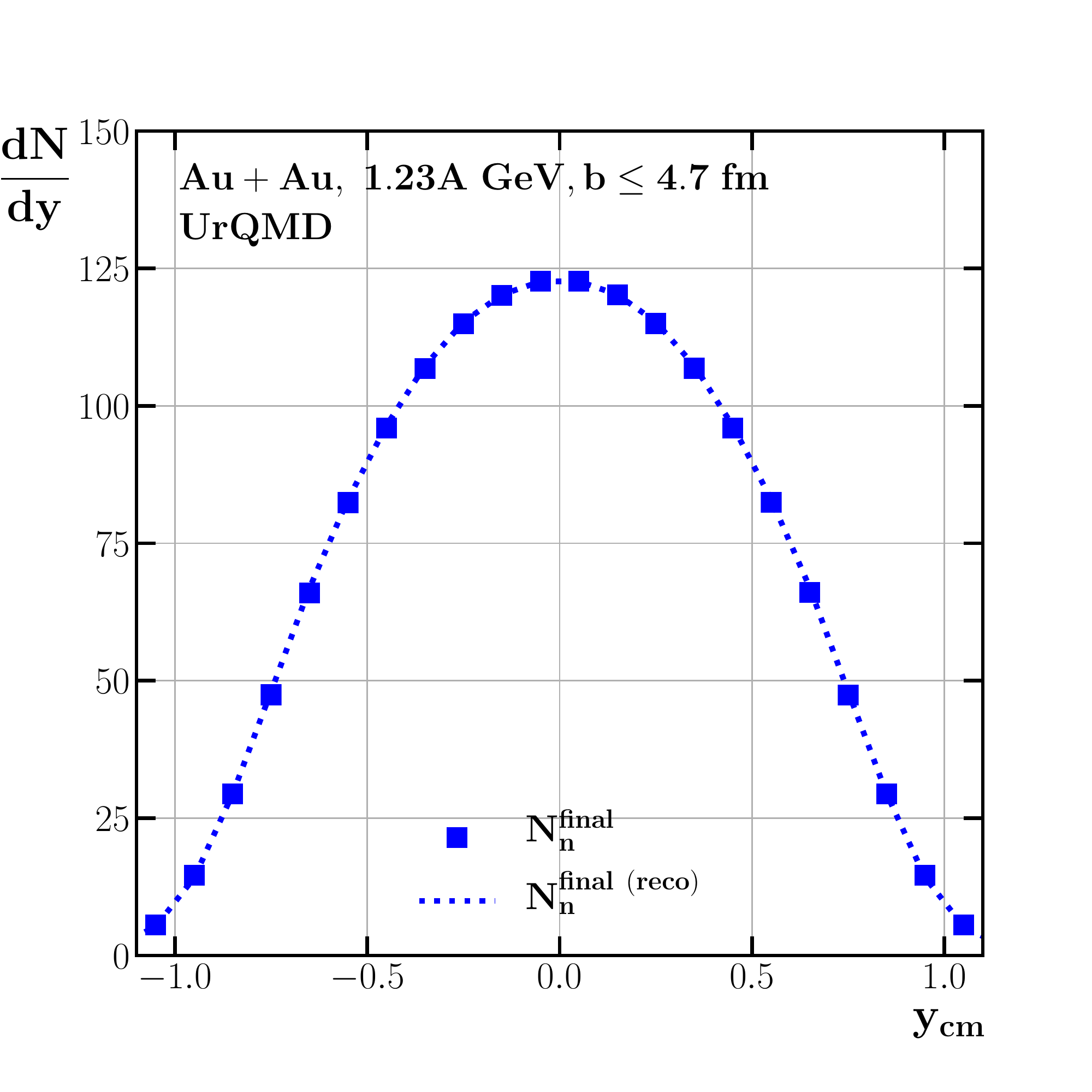}
    \caption{[Color online] Rapidity distributions of final state neutrons, reconstructed using Eq. \eqref{eq:N_n final} (dotted blue line) in comparison to the final state neutron distribution from the simulation (full blue squares), from central Au+Au reactions at $E_\mathrm{beam}=1.23A$\ GeV.}
    \label{fig:dndy_neutrons_estimator_final}
\end{figure}

As a side remark, we want to point out that the same technique can be used to extract the final state neutron rapidity distribution from the measured proton and cluster distributions and the charged pion numbers by subtracting the clusters from the primordial neutron distribution:
\begin{equation}
     \frac{{\rm d}N_n^{\rm prim\,(reco)}}{{\rm d}y} = \frac{{\rm d}N_n^{\rm final}}{{\rm d}y} + \sum\limits_{\rm cluster}N_c^n\frac{\mathrm{d}N_c^{\rm final}}{\mathrm{d}y}. 
\end{equation}
Here $N_c^n$ is the number of neutrons in each cluster $c$. By expressing everything in measurable observables this yields
\begin{widetext}
\begin{align}
    \label{eq:N_n final}
    \frac{{\rm d}N_n^{\rm final\,(reco)}}{{\rm d}y} &= \left(\frac{{\rm d}N_p^{\rm final}}{{\rm d}y} + \sum\limits_{\rm cluster}Z_c^p\frac{\mathrm{d}N_c^{\rm final}}{\mathrm{d}y}
    \right) \times \delta_{\rm iso}^{\rm prim} - \left(\sum\limits_{\rm cluster}N_c^n\frac{\mathrm{d}N_c^{\rm final}}{\mathrm{d}y}
    \right) .
\end{align}
\end{widetext}
In Fig. \ref{fig:dndy_neutrons_estimator_final} we show the distribution of final state (free) neutrons based on our estimate for the neutron distribution based on the proton distribution and Eq. \eqref{eq:N_n final} (dotted blue line) in comparison to the final state neutron distribution from the simulation (full blue squares). Also in this case the reconstruction quality is very high. 


\begin{figure}
    \centering
    \includegraphics[width=\columnwidth]{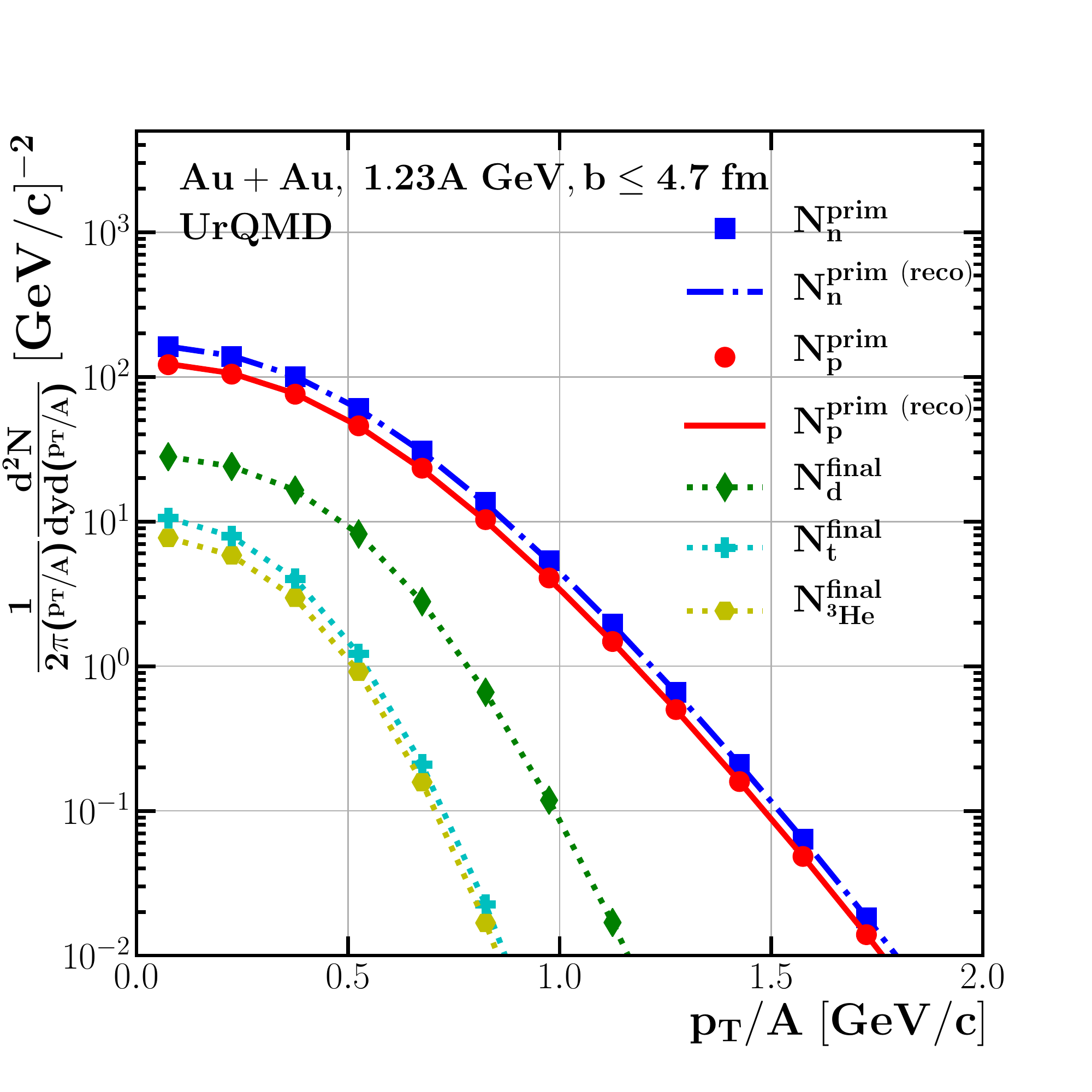}
    \caption{[Color online] Invariant distributions of final state clusters ($d$: green diamonds with dotted line, $t$: cyan pluses with dotted line, $^3$He: yellow hexagons with dotted line), the primordial proton (full red circles) and neutron (full blue squares) distribution from the simulation, and from our reconstruction (proton: solid red line, neutron: solid blue line). The calculations are at midrapidity in central Au+Au reactions at $E_\mathrm{beam}=1.23A$\ GeV.}
    \label{fig:invariant_yield}
\end{figure}

After having established a baseline by estimating the primordial neutron rapidity distribution we are now in the position to investigate the transverse momentum spectra to be able to calculate the correct $B_2$ values at low energies using both the primordial proton and primordial neutron transverse momentum distributions. To reconstruct the primordial proton and neutron transverse momentum distributions we follow the same procedure as for the rapidity distributions, but additionally scale the transverse momentum of the clusters by their respective mass number, $p_{\rm T}/A$. Like in the rapidity case this means we compare the distributions at the same (transverse or longitudinal) velocity. The primordial proton transverse momentum distribution is thus
\begin{equation}
    \frac{{\rm d}N_p^{\rm prim\,(reco)}}{p_{\rm T}{\rm d}p_{\rm T}} = \frac{{\rm d}N_p^{\rm final}}{p_{\rm T}{\rm d}p_{\rm T}} + \sum\limits_{\rm cluster}Z_c^p\frac{\mathrm{d}N_c^{\rm final}}{\frac{p_{\rm T}}{A_c}\frac{\mathrm{d}p_{\rm T}}{A_c}}
\end{equation}
and hence the primordial neutron distribution becomes
\begin{align}
    \frac{{\rm d}N_n^{\rm prim\,(reco)}}{p_{\rm T}{\rm d}p_{\rm T}} &= \left(\frac{{\rm d}N_p^{\rm final}}{p_{\rm T}{\rm d}p_{\rm T}} + \sum\limits_{\rm cluster}Z_c^p\frac{\mathrm{d}N_c^{\rm final}}{\frac{p_{\rm T}}{A_c}\frac{\mathrm{d}p_{\rm T}}{A_c}} 
    \right) \delta_{\rm iso}^{\rm prim} .
\end{align}
In Fig. \ref{fig:invariant_yield} we show the invariant distributions of final state clusters ($d$: green diamonds with dotted line, $t$: cyan pluses with dotted line, $^3$He: yellow hexagons with dotted line), the primordial proton (full red circles) and neutron (full blue squares) distribution from the simulation, and from our reconstruction (proton: solid red line, neutron: solid blue line). The calculations are at midrapidity ($|y|\leq0.5$) in central Au+Au reactions at $E_\mathrm{beam}=1.23A$\ GeV.
\begin{figure}
    \centering
    \includegraphics[width=\columnwidth]{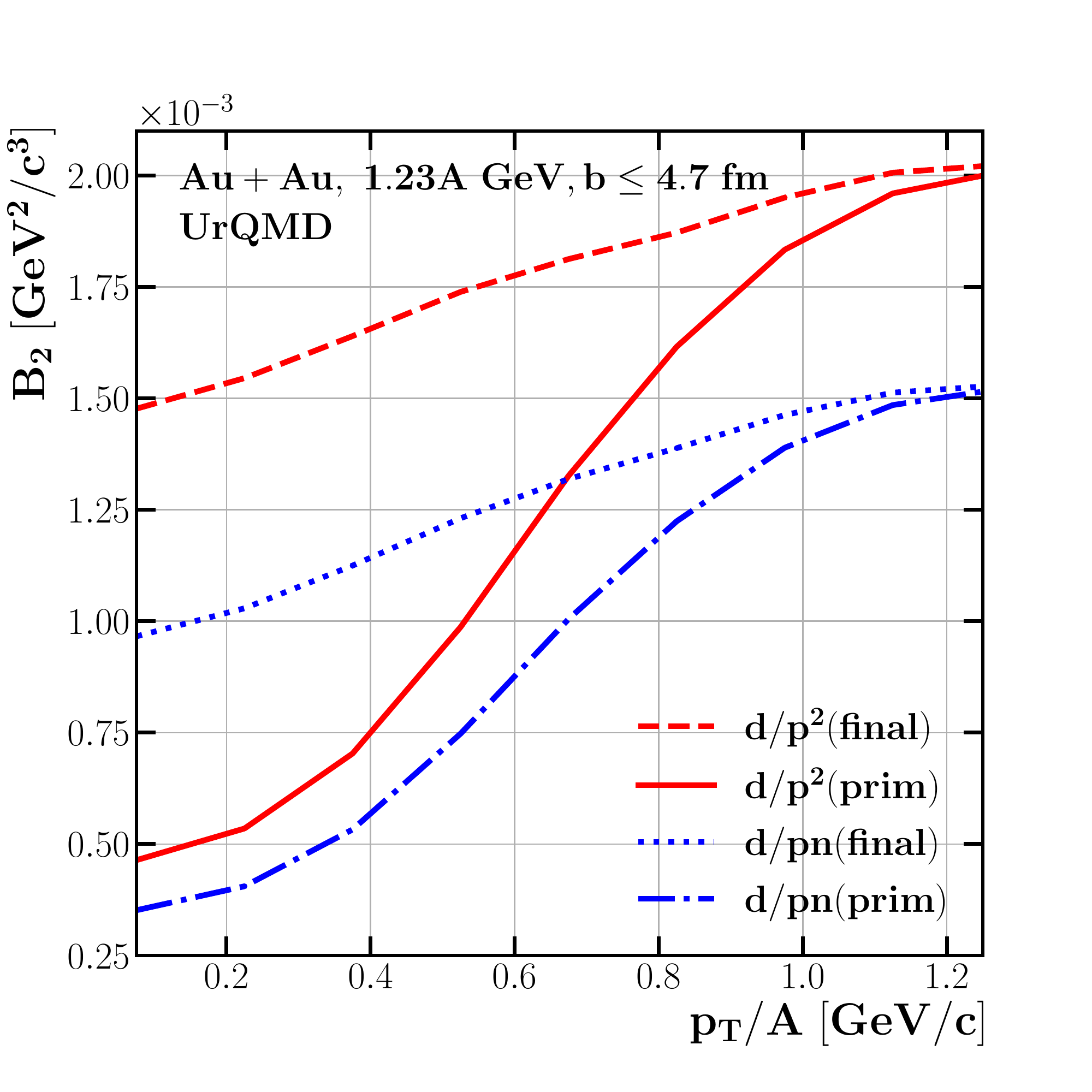}
    \caption{[Color online] $B_2$ as a function of scaled transverse momentum for central Au+Au reactions at $E_\mathrm{beam}=1.23A$ GeV. The dashed red line shows $B_2$ using the squared final state proton yield, the solid red line shows the result using the primordial protons squared, the dotted blue line shows the $B_2$ using the final state protons and neutrons and the dash-dotted blue line depicts the fully corrected result using the primordial protons and neutrons for the ratio. The calculations are at midrapidity in central Au+Au reactions at $E_\mathrm{beam}=1.23A$\ GeV.} 
    \label{fig:B2_vs_pT}
\end{figure}
We observe that also the transverse momentum spectra of the primordial protons and neutrons are reconstructed with high precision. This allows us to use the reconstructed primordial transverse momentum distributions to calculate the corrected $B_2$ values.

\subsection{Corrected $B_2$ estimates}
The invariant distribution of the primordial protons and neutrons and the final state deuterons now serve as input to calculate $B_2$ as a function of transverse momentum per nucleon $p_\mathrm{T}/A$. The $B_2$ parameter can be obtained by taking the ratios of the invariant yields, i.e. $B_2$ equals $E_{d}\frac{{\rm d}^3N_{d}}{{\rm d} p_{d}^3}$ at $p_{d}/2$ divided by the product of $\left(E_p\frac{{\rm d}^3N_p}{{\rm d} p_p^3}\right) \cdot \left(E_n\frac{{\rm d}^3N_n}{{\rm d} p_n^3} \right)$ from the primordial state before clustering. Also the $B_3$ can be calculated using the neutron distributions instead of the cube of the proton distributions, i.e. $B_3$ equals $E_{t}\frac{{\rm d}^3N_{t}}{{\rm d} p_{t}^3}$ at $p_{t}/3$ divided by the product of $\left(E_p\frac{{\rm d}^3N_p}{{\rm d} p_p^3}\right)\cdot \left(E_n\frac{{\rm d}^3N_n}{{\rm d} p_n^3}\right)^2$ from the primordial state before clustering. Also the $B_3$ of the $^3$He can be calculated in this way, i.e. by dividing $E_{^3\mathrm{He}}\frac{{\rm d}^3N_{^3\mathrm{He}}}{{\rm d} p_{^3\mathrm{He}}^3}$ at $p_{^3\mathrm{He}}/3$ by the product of $\left(E_p\frac{{\rm d}^3N_p}{{\rm d} p_p^3}\right)^2 \cdot \left(E_n\frac{{\rm d}^3N_n}{{\rm d} p_n^3}\right)$ from the primordial state before clustering. 
\begin{figure}[t!]
    \centering
    \includegraphics[width=\columnwidth]{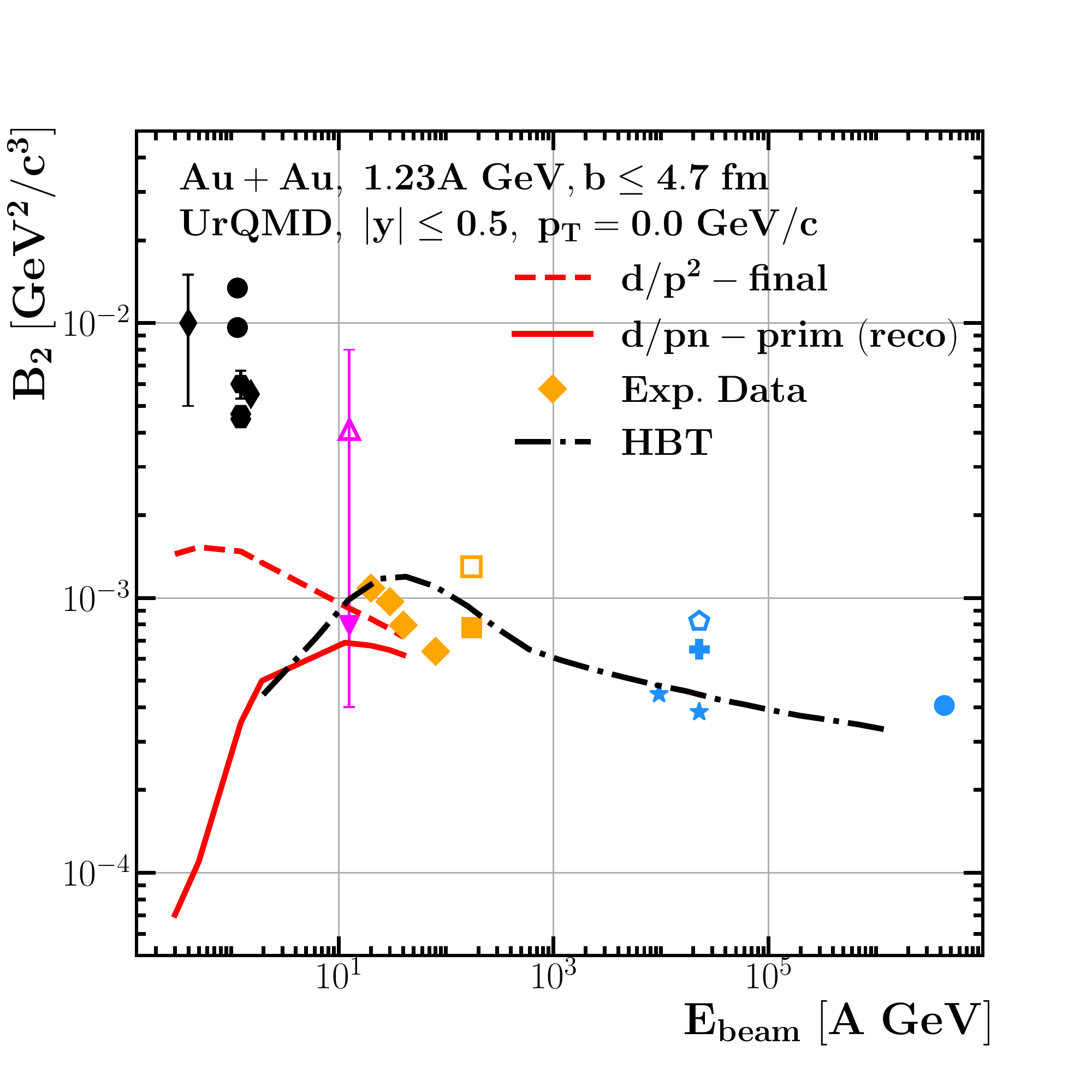}
    \caption{[Color online] $B_2$ extracted at $p_\mathrm{T}/A=0$ GeV at midrapidity $|y|\leq0.5$ as a function of beam energy for central Au+Au reactions. The dashed red line shows the calculation using the final state proton yields (i.e. uncorrected), while the solid red line shows the corrected $B_2$ values using our estimates for the distribution before clustering and taking into account the estimated neutron yields. The experimental data \cite{FOPI-B2,EOS:1994jzn,HADES-B2,E864:1999dqo,E802:1999hit,E864:2000auv,E877:1999qpr,NA52NEWMASS:1997gjo,Bearden:2002ta,NA49:2016qvu,NA49:2004mrq,NA49:2011blr,NA49:2000kgx} is shown by symbols. The dash-dotted black line shows the volume extracted from HBT results from STAR \cite{STAR:2014shf}.}
    \label{fig:B2_correct}
\end{figure}
\begin{figure}[t!]
    \centering
    \includegraphics[width=\columnwidth]{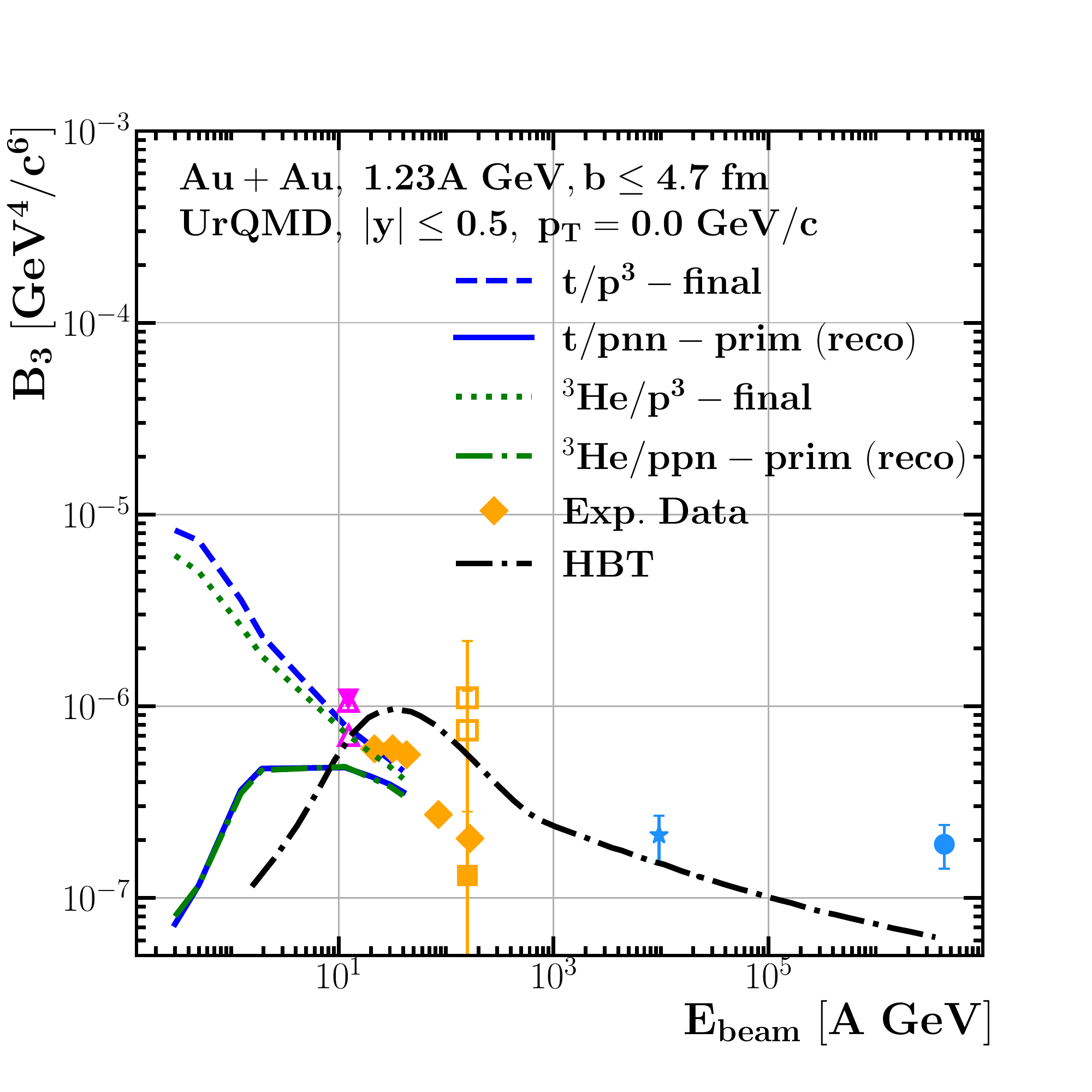}
    \caption{[Color online] $B_3$ extracted at $p_\mathrm{T}/A=0$ GeV at midrapidity $|y|\leq0.5$ as a function of beam energy for central Au+Au reactions. The dashed blue ($t$) and the dotted green ($^3$He) lines show the calculations using the final state yields (i.e. uncorrected), while the solid blue ($t$) and solid green ($^3$He) lines show the corrected $B_3$ values using our estimates for the distribution before clustering and taking into account the estimated neutron yields. The experimental data \cite{FOPI-B2,EOS:1994jzn,HADES-B2,E864:1999dqo,E802:1999hit,E864:2000auv,E877:1999qpr,NA52NEWMASS:1997gjo,Bearden:2002ta,NA49:2016qvu,NA49:2004mrq,NA49:2011blr,NA49:2000kgx} is shown by symbols. The dash-dotted black line shows the volume extracted from HBT results from STAR \cite{STAR:2014shf}.}
    \label{fig:B3_correct}
\end{figure}

In order to show the impact of the effects discussed above, Fig. \ref{fig:B2_vs_pT} illustrates the calculated $B_2$ values as a function of the transverse momentum per nucleon by taking the ratio of the deuteron distribution and final state proton distribution squared (dashed red line), the primordial proton distribution squared (solid red line), the final proton and reconstructed neutron distribution (dotted blue line) and the reconstructed primordial proton and neutron distribution (dash-dotted blue line) for central Au+Au collisions at a kinetic beam energy of $1.23A$\ GeV from UrQMD. This clarifies the fundamental distinctions between final state and primordial protons (or neutrons). 
\begin{enumerate}
    \item Let us first discuss the impact of the neutrons in the final state: the $B_2$ from the final state protons squared (dashed red line) is about $1.2-1.5$ times the $B_2$ from the product of final state protons and neutrons (dotted blue line). This scaling is visible over the full range of transverse momenta per nucleon, i.e. it is independent of $p_\mathrm{T}/A$. In case of the primordial state, the ratio between $B_2$ from the primordial proton squared (solid red line) and the product of primordial state of proton and neutron (dash-dotted blue line) is similarly around $1.2-1.5$ and also not $p_{\rm T}/A$ dependent. These ratios align well with the isospin asymmetry factor from the initial gold nuclei $N_{\rm Au}/Z_{\rm Au} = 1.49$ and the isospin asymmetry at the primordial stage $\delta_{\rm iso}^{\rm prim}=1.32$ as expected.

    \item Then let us discuss about the impact of using primordial protons (neutrons): by comparing $B_2$ from the final state protons squared (dashed red line) with $B_2$ from the primordial protons squared (solid red line) we observe a huge reduction of $B_2$ on the order of $\approx3$ at low transverse momenta, but nearly no reduction at high transverse momenta. This is because the primordial protons (and neutrons) are more likely to coalesce into deuterons (and other clusters) at low $p_{\rm T}/A$ and less likely at high $p_{\rm T}/A$ (see Fig. \ref{fig:invariant_yield}) and hence the effect is transverse momentum dependent. Consequently, for larger $p_{\rm T}/A$, a minor difference is expected between primordial protons (and neutrons) and final state protons (and neutrons) which is indeed the case as demonstrated by the convergence of the solid red (dash-dotted blue) lines towards the dashed red (dotted blue) lines.
\end{enumerate}  The impact of the full correction, i.e. $B_2$ from reconstructed primordial protons and neutrons (dash-dotted blue line) is greatest at $p_{\rm T}/A=0$ GeV in comparison to the final proton distribution squared (dotted red line). The $B_2$ at zero transverse momentum is thus reduced by a factor of $\approx 4$ taking into account both corrections.

Finally, we analyze the energy dependence of $B_2$ and $B_3$ and compare to the experimental data. Here we choose $p_{\rm T}/A=0$ GeV in line with previous studies by the HADES collaboration. In Fig. \ref{fig:B2_correct} we show the $B_2$ as a function of energy for central Au+Au reactions. The dashed red line shows the calculation using the final state proton yields (i.e. uncorrected), while the solid red line shows the corrected $B_2$ values using our estimates for the primordial distributions before clustering and taking into account the estimated neutron yields (our data is summarized in Tab. \ref{tab:B_2} in the appendix). The experimental data \cite{FOPI-B2,EOS:1994jzn,HADES-B2,E864:1999dqo,E802:1999hit,E864:2000auv,E877:1999qpr,NA52NEWMASS:1997gjo,Bearden:2002ta,NA49:2016qvu,NA49:2004mrq,NA49:2011blr,NA49:2000kgx} is shown by symbols. The dash-dotted black line shows the volume extracted from HBT results from STAR \cite{STAR:2014shf}. We observe that the correction is substantial at low energies ($\sqrt{s_{\rm NN}}< 6$ GeV) and becomes small towards higher collision energies.

In Fig. \ref{fig:B3_correct} we show the $B_3$ as a function of energy for central Au+Au reactions. The dashed blue ($t$) and the dotted green ($^3$He) lines show the calculations using the final state proton yields (i.e. uncorrected), while solid blue ($t$) and the solid green ($^3$He) lines show the corrected $B_3$ values using our estimates for the distribution before clustering and taking into account the estimated neutron yields (our data is summarized in Tab. \ref{tab:B_3} in the appendix). The experimental data \cite{FOPI-B2,EOS:1994jzn,HADES-B2,E864:1999dqo,E802:1999hit,E864:2000auv,E877:1999qpr,NA52NEWMASS:1997gjo,Bearden:2002ta,NA49:2016qvu,NA49:2004mrq,NA49:2011blr,NA49:2000kgx} is shown by symbols. The dash-dotted black line shows the volume extracted from HBT results from STAR \cite{STAR:2014shf}. We observe that the correction is substantial at low energies ($\sqrt{s_{\rm NN}}< 6$ GeV) and also becomes small towards higher collision energies. One should also note that the splitting between triton and $^3$He vanishes, if $B_3$ is calculated with the correct primordial neutron yields.

In contrast to the usual procedure, which suggests that the $B_2$ and $B_3$ values increase towards lower energies, the corrected procedure leads to a decrease of $B_2$ and $B_3$ with decreasing energy. This brings the $B_2$ and $B_3$ values back in line with the estimates from HBT measurements.

\section{Conclusion}
We have employed the Ultra-relativistic Quantum-Molecular-Dynamics model to study proton, neutron and cluster yields at low energies. We showed that the apparent observation of an increasing $B_2$ and $B_3$ towards low energies is due to the use of final state proton distributions and not primordial (i.e. before coalescence as required by the coalescence equation) distributions. In addition, we showed that the use of the proton distribution as a proxy for the neutron distribution is not justified. Here we suggest to estimate the neutron yield from the measured proton, cluster and pion yields. Both effects reduce the $B_2$ and $B_3$ values substantially and bring them in line with the HBT measurements. In general we strongly suggest to use the corrected $B_A$ values at energies below $\sqrt{s_{\rm NN}}< 6$ GeV.

\begin{acknowledgements}
This article is part of a project that has received funding from the European Union’s Horizon 2020 research and innovation program under grant agreement STRONG – 2020 - No 824093. J.S. thanks the Samson AG for funding. Computational resources were provided by the Center for Scientific Computing (CSC) of the Goethe University and the ``Green Cube" at GSI, Darmstadt. This project was supported by the DAAD (PPP Thailand). This research has received funding support from the NSRF via the Program Management Unit for Human Resources \& Institutional Development, Research and Innovation [grant number B16F640076].
\end{acknowledgements}

\section*{Appendix}
Here we show the numerical values $B_2$ and $B_3$. The $B_2$ values are compared between the usual procedure ($d/p^2_{\rm final}$ using the final state yields) and the corrected procedure using the primordial proton and neutron yields ($d/p_{\rm prim}n_{\rm prim}$). Similarly for $B_3$.

\begin{table}[h!]
    \centering 
    \renewcommand{\arraystretch}{1.5} 
    \begin{tabular}{c||c|c}
         \multirow{2}{*}{E$_\mathrm{beam}$ [$A$\ GeV]} & \multicolumn{2}{c}{$B_2$ [$\times10^{-4}$GeV$^2$/c$^3$]} \\ \cline{2-3}
         & $d/p^2_{\rm final}$ & $d/p_{\rm prim}n_{\rm prim}$ \\
         \hline
         \hline
        0.3 & 14.44 & 0.70 \\ \hline
        0.5 & 15.31 & 1.10 \\ \hline
        1.23 & 14.77 & 3.52 \\ \hline
        1.93 & 13.44 & 5.00 \\ \hline
        11.45 & 9.34 & 6.88 \\ \hline
        20 & 8.39 & 6.71 \\ \hline
        30 & 7.72 & 6.46 \\ \hline
        40 & 7.22 & 6.21 \\ \hline
    \end{tabular}
    \caption{Calculated $B_2$ values using final state protons and primordial protons and neutrons at $p_{\rm T}/A=0.0$ GeV at midrapidity $|y|\leq0.5$ from central Au+Au collisions at kinetic beam energies from $0.3$ to $40A$\ GeV.}
    \label{tab:B_2}
\end{table}

\begin{table}[h!]
    \centering  
    \renewcommand{\arraystretch}{1.5} 
    \begin{tabular}{c||c|c|c|c}
         \multirow{2}{*}{E$_\mathrm{beam}$ [$A$\ GeV]} & \multicolumn{2}{c|}{$B_3^t$ [$\times10^{-7}$GeV$^4$/c$^6$]} & \multicolumn{2}{c}{$B_3^{^3\mathrm{He}}$ [$\times10^{-7}$GeV$^4$/c$^6$]} \\ \cline{2-5}
         & $t/p^3_{\rm final}$ & $t/p_{\rm prim}n_{\rm prim}^2$ & $^3$He$/p^3_{\rm final}$ & $^3$He$/p_{\rm prim}^2n_{\rm prim}$ \\
         \hline
         \hline
         0.3 & 82.75 & 0.73 & 61.17 & 0.80 \\ \hline
        0.5 & 72.93 & 1.16 & 50.16 & 1.16 \\ \hline
        1.23 & 36.00 & 3.64 & 26.24 & 3.50 \\ \hline
        1.93 & 23.35 & 4.72 & 18.25 & 4.63 \\ \hline
        11.45 & 7.94 & 4.78 & 7.29 & 4.83 \\ \hline
        20 & 6.22 & 4.29 & 5.79 & 4.22 \\ \hline
        30 & 5.25 & 3.91 & 4.83 & 3.81 \\ \hline
        40 & 4.58 & 3.56 & 4.18 & 3.41 \\ \hline
    \end{tabular}
    \caption{Calculated $B_3$ values using final state protons and primordial state protons and neutrons at $p_{\rm T}/A=0.0$ GeV at midrapidity $|y|\leq0.5$ from central Au+Au collisions at kinetic beam energies from $0.3$ to $40 A$\ GeV.}
    \label{tab:B_3}
\end{table}



\end{document}